\documentclass[conference]{IEEEtran}

\usepackage{graphicx}
\usepackage{amsfonts}
\usepackage[cmex10]{amsmath}
\usepackage{cite}
\usepackage{algorithmic}
\usepackage[center]{caption}
\usepackage{epsfig}
\usepackage{latexsym}
\usepackage{epstopdf}
\usepackage{verbatim}
\usepackage{amsthm}
\usepackage{balance}
\usepackage{subfig}
\usepackage{epstopdf}

\usepackage[linesnumbered,ruled,vlined]{algorithm2e}
 
\graphicspath{{./figures/}}

\usepackage{bm}
\newcommand{\vect}[1]{\boldsymbol{\mathbf{#1}}}

\begin{document}
\title{A Probabilistic MAC for Cognitive Radio Systems with Energy Harvesting Nodes}

\author{\large Ramy E. Ali$^1$, Fadel F. Digham$^2$, Karim G. Seddik$^3$, Mohammed Nafie$^1$, Amr El-Keyi$^1$ and Zhu Han$^4$ \\ [.1in] 
\small  
\begin{tabular}{c}
$^1$Wireless Intelligent
Networks Center (WINC), Nile University, Egypt.\\
$^2$National Telecom Regulatory Authority (NTRA), Egypt. \\
$^3$Electronics Engineering Department, American University in Cairo, AUC Avenue, New Cairo 11835, Egypt.\\

$^4$ Computer Engineering Department, University of Houston, Houston, TX, USA \\
email: ramy.essam@nileu.edu.eg, fadel.digham@ieee.org, kseddik@aucegypt.edu, \{mnafie, aelkeyi\}@nileuniversity.edu.eg, zhan2@uh.edu
\end{tabular} }

\maketitle
\begin{abstract}
In this paper, we consider a cognitive radio (CR) system where the secondary user (SU) harvests energy from both the nature resources and the primary user (PU) radio frequency (RF) signal.  We propose an energy-based probabilistic access scheme in which SU probabilistically accesses and senses the primary channel. The decision is based on the available energy and the PU's activity.  We investigate the problem of maximizing the SU's success rate provided that the PU average quality of service (QoS) constraint is satisfied. We also assume  multi-packet reception (MPR) capability and sensing errors under a Rayleigh fading channel. Numerical
results show the effectiveness of the proposed probabilistic access scheme.
\end{abstract}

\IEEEpeerreviewmaketitle

\section{Introduction}
\makeatletter{\renewcommand*{\@makefnmark}{}
\footnotetext{\hrule \vspace{0.05in} This work was supported by a grant from the Egyptian National Telecommunications Regulatory Authority (NTRA).

Mohamed Nafie is also affiliated with the EECE
Dept., Faculty of Engineering, Cairo University.
}\makeatother }
Cognitive radio (CR) is a promising technology in which the secondary user (SU) can opportunistically access the licensed spectrum held by the primary user (PU) provided that the PU minimum quality of service (QoS) constraint is maintained, hence the spectral efficiency can be improved \cite{akyildiz2006next}.

 Energy harvesting (EH) provides a free source of energy for the wireless nodes \cite{mateu2005review}. In nature EH (NEH), wireless devices can utilize the free renewable energy sources such as solar energy to improve the energy efficiency. The major drawback of the NEH is that the wireless devices cannot depend only on the nature energy due to its randomness, dependence on the environmental changes, and the location \cite{saleh2013department}. Radio frequency (RF) based EH is an energy recycling process through which a low-power device can harvest energy from the ambient RF signals, with a certain efficiency, and reuse it again for transmission \cite{lu2014dynamic}. The major limiting factor of the RF EH is that RF harvested energy decays with the increase in distance between the RF source and RF harvester.

EH CR networks have got a lot of attention recently. In \cite{sultan2012sensing}, a CR system is considered where the SU harvests energy from nature resources. The secondary transmitter (ST) may sense the channel and decide to access if it is sensed idle or may remain idle. This decision is based on the available energy and the ST's belief about the PU's activity. Assuming a collision, maximizing the SU reward is addressed using a dynamic programming framework. In \cite{lee2013opportunistic}, an RF-powered CR system is modeled using a stochastic geometry approach. Each PU is associated with a guard zone to protect it from the SU's interference. STs can harvest RF energy if they are in the harvesting zone of a PU. The problem of maximizing the SU throughput is studied under PU and SU minimum QoS constraints. In \cite{chung2014spectrum}, a CR system is considered where the ST has a rechargeable infinite battery and harvests nature energy. The ST accesses the PU's spectrum if it is sensed idle. The ST adjusts the sensing threshold and time to maximize its throughput subject to PU QoS.

In this paper, we propose a probabilistic scheme in a mixed EH CR system with one PU and one SU. NEH is limited by the environment and RF EH is limited by the distance between
the RF source and the harvester. Mixed EH means that the ST harvests energy from both nature resources and the primary transmitter (PT) RF signal, so the energy efficiency can be enhanced. The ST has two wireless interfaces one for the RF EH and other for transmission. Assuming an interference channel model with multi-packet reception capability \cite{ghez1988stability}, the SU probabilistically accesses the channel provided that the PU QoS constraint is satisfied. Based on the available energy, the ST may remain idle, access the spectrum without sensing, or sense the channel $\&$ access if it is sensed idle. The ST adjusts the sensing parameters to maximize its success rate. Moreover, we consider the fading effects on the sensing unlike the assumptions made in \cite{sultan2012sensing} and \cite{chung2014spectrum}, where they assume a fixed channel realization. Since sensing requires energy, the SU may access the channel without sensing to save this energy and to dedicate more time for transmission. On the other hand, accessing the channel without sensing may increase the SU's outage probability and violate the PU's QoS constraint, hence we address this trade-off. 

The remainder of this paper is organized as follows, In Section \ref{System Model}, we present the system model. In Section \ref{Success Rate Analysis}, we address the problem of maximizing the SU success rate. The simulation results are presented in Section \ref{Simulation Results}. Finally, concluding remarks are drawn in Section \ref{Conclusions}.
 
\section{System Model}
\label{System Model}

\begin{figure}[h]
  \centering
\includegraphics[width=0.6\linewidth,height=.145\textheight]{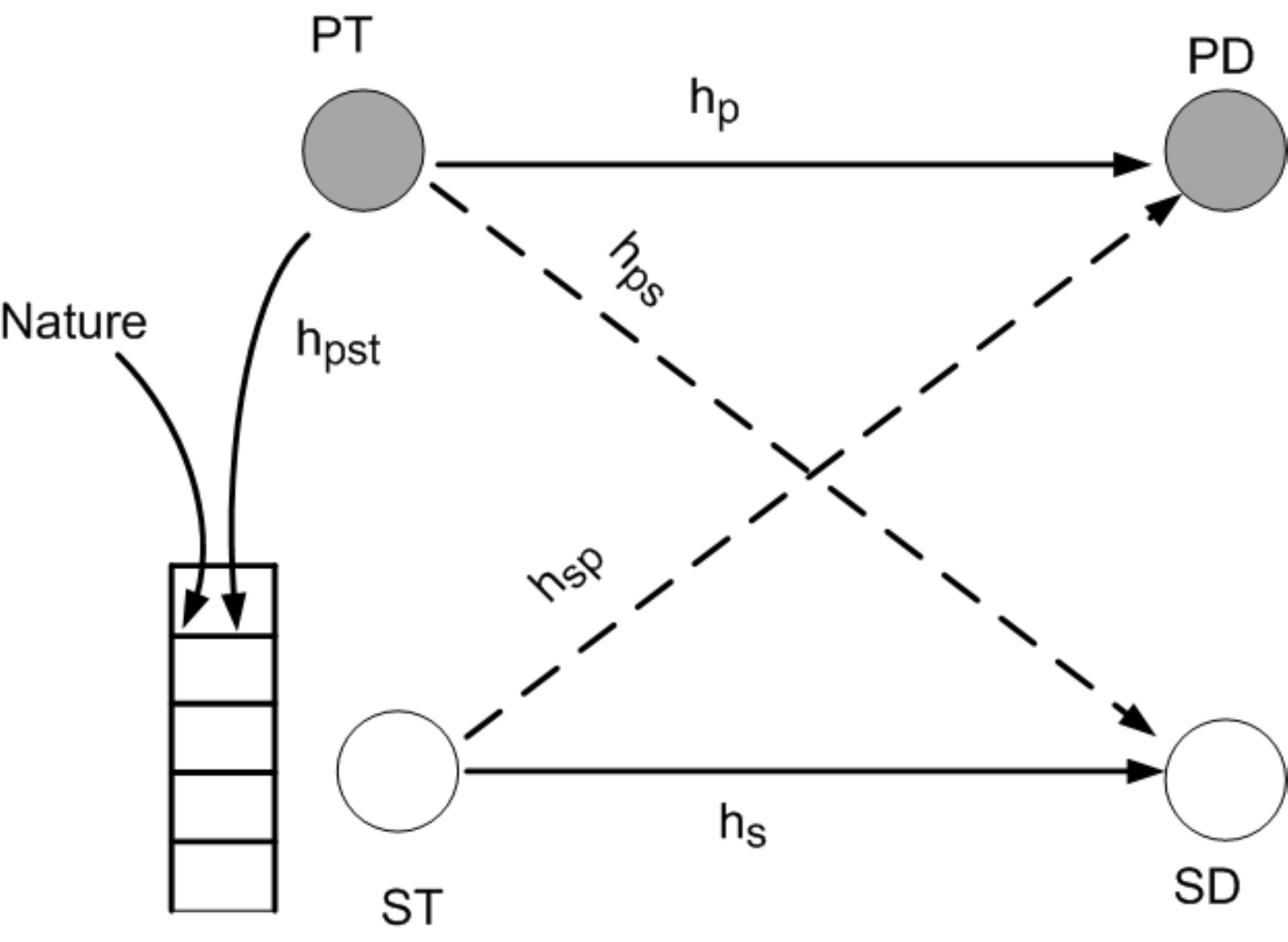} 
\caption{\small System Model \label{fig 1}}
\end{figure}

We consider a slotted time CR system with one PU and one SU as shown in Fig. \ref{fig 1}. The SU has a rechargeable battery which is modeled as a queue, $Q_e$, with a limited capacity, $N_{\rm max}$. The ST can harvest energy from both nature resources and ambient RF signals. We also assume Rayleigh flat fading channels with constant channel gain during a time slot, $T$. The channel coefficient accounts for the fading and path-loss effects. We denote the channel coefficient of the PU direct link by $h_{\rm p}$, the link PT-ST by $h_{\rm pst}$, the link PT-SD by $h_{\rm ps}$, the link ST-SD by $h_{\rm s}$ and the link ST-PD by $h_{\rm{sp}}$. The channel gain of a link $i$-$j$, $|h_{ij}|^2$ follows an exponential distribution with mean $\sigma_{ij}=\tilde{\sigma}_{ij}/ d_{ij}^2$, where $d_{ij}$ is the length of this link and $\tilde{\sigma}_{ij}$ is the mean of the fading coefficient. We also assume additive white Gaussian noise (AWGN) with variance $\sigma_n^2$.

\subsection{Nature Energy Harvesting}
\label {Nature}
If we divide the slot time $T$ into small intervals each of length $\delta$, then we have $n=T/ \delta$ mini-slots. We assume that the probability of receiving one energy packet within a mini-slot is $p$. Then the number of harvested nature energy packets within $T$ is a binomial random variable (RV) with parameters $n$ and $p$. If we assume that $n$ is very large and $p$ is very small, this RV can be approximated as a Poisson RV with a parameter $\lambda_e T= n p$ \cite[Ch.\ 4, p. 113]{papoulis2002probability}, where $\lambda_e$ is the energy arrival rate (energy packets / sec).
 The Poisson distribution captures the randomness of the nature energy arrivals and the possibility of harvesting any number of energy packets in the time slot \cite{xu2014throughput}. Define $p_N(n)$ as the probability of harvesting $n$ packets from the nature during time $T$ as follows, 
\begin{small}
{
\begin{equation}
p_N(n)= \frac{(\lambda_e T)^n \exp (-\lambda_e T)}{n!}, \ \ n=0,1,2,\cdots. 
\end{equation}
}
\end{small}
\vspace{-18pt}
\subsection{RF Energy Harvesting} 
\label{RF}

The ST can harvest RF energy from the PU transmissions. The efficiency of this process $\eta$  depends on the RF-to-DC conversion circuit. For commercially available RF energy harvesters, $\eta$ ranges from $0.5$ to $0.7$ \cite{rf}. The number of harvested packets can be approximated as follows \cite{zhou2012wireless},
\begin{small}
{
\begin{equation}
\mathcal{R}_e=\lfloor \frac{\eta P_{\rm{p}} |h_{\rm pst}|^2 T}{E_{\rm u}}\rfloor ,
\end{equation}
}
\end{small}\!where $E_{\rm u}$ is the amount of Joules per energy packet. The distribution of the RV $\mathcal{R}_e$ can be derived as follows 
\begin{small}
{
\begin{equation}
\begin{split}
& p_{\mathcal{R}_e}(r)=\Pr \{\mathcal{R}_e=r\}= \Pr \{r \leq \frac{\eta P_{\rm p} |h_{\rm pst}|^2 T}{E_{\rm u}} < (r+1) \}\\
&=\exp \left( \frac{-r E_{\rm u}}{\sigma_{\rm{pst}} \eta P_{\rm p} T}   \right)- \exp \left( \frac{-(r+1) E_{\rm u}}{\sigma_{\rm pst} \eta P_{\rm p} T}   \right) \ \ r=0,1,2, \cdots.
\end{split}
\end{equation}
}
\end{small}
\subsection{Sensing and energy queue dynamics}
\label{Sensing}\!The ST consumes some energy in the sensing process depending on the sensing time $\tau$ and it is given by \cite{sultan2012sensing}
\begin{equation}
E_{\rm s} =n_{\rm s} \ e_{\rm proc} ,
\end{equation}
where $e_{\rm proc}$ is the energy required to process one sample, and $n_s$ is the number of collected samples during the sensing interval and given by $n_s=f_s \tau$, where $f_s$ is the sampling frequency of the sensing process. The number of energy packets used for sensing and transmission $\mathcal{N}_s$ and $\mathcal{N}_t$ are given by $\mathcal{N}_s(\tau) =\lceil \frac{E_{\rm s}}{E_{\rm u}} \rceil$ and $\mathcal{N}_t=\lceil \frac{E_t}{E_u} \rceil$ respectively, where $E_t$ is the transmit energy of the ST and assumed to be constant.


The SU is assumed to have a packet of length $b_{\rm s}$ at each time slot and can access the PU's channel provided that the PU QoS constraint is satisfied. The PT has a packet of length $b_{\rm p}$ at each slot with probability $\rho$, which is the prior probability that the PU's channel is occupied. The ST selects its actions based on the available energy in the battery, $\mathcal{Q}_e$ as follows:
\begin{enumerate}
\item $0< \mathcal{Q}_e < \mathcal{N}_t$, the ST has to remain idle.
\item If $\mathcal{N}_t \leq \mathcal{Q}_e < \mathcal{N}_t+\mathcal{N}_s$, the ST may access the channel without sensing with probability $\alpha_{q_e}$ or remain idle with probability $1-\alpha_{q_e}$.
\item If $\mathcal{N}_t+\mathcal{N}_s \leq \mathcal{Q}_e \leq N_{\rm max} $, the ST may access the channel without sensing with probability $\beta_{1_{q_e}}$, sense the channel with probability $\beta_{2_{q_e}}$ or remain idle with probability $1-(\beta_{1_{q_e}}+\beta_{2_{q_e}})$.
 \end{enumerate}

 The ST adopts the energy detection sensing technique to track the PU activity. For a given realization of the channel $h_{\rm pst}$, the probability of detection is given by \cite{digham2007energy}
\begin{equation}
P_d= Q_m(\sqrt{2 \gamma},\sqrt{\lambda}) ,
\end{equation}
and the false alarm probability is given by
\begin{small}{
\begin{equation}
\label{false alarm}
P_F=\frac{\Gamma (m, \lambda/2)}{\Gamma (m)} ,
\end{equation}
}
\end{small}
\\ where $\lambda$ is the detection threshold, $Q_m(., .)$ is the generalized Marcum Q-function, $\Gamma (.)$ is the complete gamma function, $\Gamma (.,.)$ is the incomplete gamma function, $\gamma$ is the signal to noise ratio (SNR) and $m$ is the time bandwidth product, which is given by $m=\tau W$ and it is assumed to be integer. By taking the average of the $P_d$ over $|h_{\rm pst}|^2$, the average probability of detection is given by \cite{digham2007energy}
\begin{small}
\begin{equation}
\label{detection}
\begin{split}
& P_D=e^{-\frac{\lambda}{2}} \sum_{k=0}^{m-2} \frac{1}{k!} \left(\frac{\lambda}{2} \right) ^k+\\
& \left(  \frac{1+\overline \gamma}{\overline {\gamma}} \right)^{m-1} \left( e^{-\frac{\lambda}{2(1+\overline {\gamma})}}-e^{-\frac{\lambda}{2}} \sum_{k=0}^{m-2} \frac{1}{k!} \left(\frac{\lambda \overline \gamma}{2(1+\overline \gamma)}\right)^k  \right) , 
\end{split}
\end{equation}
\end{small}
where $\overline \gamma$ is the average SNR and is given by $P_{\rm p}\ \sigma_{\rm pst}/ \sigma_n^2$. \\ 
\label{Steady State Probabilities}
Define $\vect{\pi}$ as the steady state probability distribution of the energy queue
length, $\vect{\pi}= [ \pi_0, \pi_1, \dotsc, \pi_{N_{\rm max}} ]$, where $\pi_{q_e}$ is the probability that the ST has $q_e$ packets in its energy queue. The energy queue dynamics can be represented by a Markov chain. We can get $\vect{\pi}$ by solving the following two equations
\begin{small}
{
\begin{eqnarray}
\label{steady state equation}
\vect{\pi} P= \vect \pi ,\\
\sum_{q_e=0}^{N_{\rm max}} \pi_{q_e} = 1,
\end{eqnarray}
}
\end{small}\!where $P$ is the transition matrix of the energy queue. $P_{ij}$ is the transition probability from state $i$ to state $j$ (the derivation of $P$ is in the \textbf{Appendix}). The sensing probability is given by
\begin{equation}
p_S=\sum_{q_e=\mathcal{N}_t+\mathcal{N}_s}^{N_{\rm max}} \pi_{q_e} \beta_{2_{q_e}} \cdot
\end{equation}
The expected sensing time is given by $\overline\tau= p_S\tau$ and the probability of access without sensing $p_A$ is given by
\begin{equation}
p_{\mathcal{A}}=\sum_{q_e=\mathcal{N}_t}^{\mathcal{N}_t+\mathcal{N}_s-1} \pi_{q_e} \alpha_{q_e}+\sum_{q_e=\mathcal{N}_t+\mathcal{N}_s}^{N_{\rm max}} \pi_{q_e} \beta_{1_{q_e}} \cdot
\end{equation}
\vspace{-10pt}

\section{Problem Formulation}
\label{Success Rate Analysis}
In this section, we derive the success rates of the PU and the SU denoted by $\mu_p$ and $\mu_{s}$ respectively. Then we address the problem of maximizing $\mu_s$ provided that the PU average QoS is maintained. The success rate is the probability of no outage. Let $\overline{P}_{\rm nsu, out}$ be the probability of no PU outage in case that the ST does not access the channel which can be expressed as \footnote{The bar sign represents the complement probability, i.e, $\overline{p}=1-p.$}
\begin{equation}
\begin{split}
 &\overline{P}^{\rm p}_{\rm nsu, out}= \Pr\{\mathcal{R}_{\rm p}< \log_2(1+\frac{P_{\rm p} |h_{\rm p}|^2}{\sigma_{\rm n}^2})\} \\
 &= \exp \left( \frac{-(2^{\mathcal{R}_{\rm p}}-1) \sigma_{\rm n}^2}{P_{\rm p} \sigma_{\rm p}}  \right) \cdot
\end{split}
\end{equation}
where $\mathcal{R}_{\rm p}=b_{\rm p}/T/W \ \rm{bits/ sec/ Hz}$ is the PU spectral efficiency. The transmission power of the SU in case of access without sensing is $E_t/T$. Hence, we can get the probability of no PU outage in this case $\overline{P}^{\rm p}_{\rm ws, out}$ as follows
\begin{small}
\begin{equation}
\label{outage}
\begin{split}
& \overline{P}^{\rm p}_{\rm ws, out}= \Pr\{\mathcal{R}_{\rm p}< \log_2(1+\frac{P_{\rm p} \vert \ h_{\rm p}|^2}{ 
(E_t/T)|h_{\rm sp}|^2+\sigma_n^2})\}\\
& =\mathbb{E}_{\rm{|h_{sp}|^2}} \left[  \Pr\left\{\frac{(2^{\mathcal{R}_{\rm p}}-1)}{P_{\rm p}}(\frac{E_t}{T}
 |h_{\rm sp}|^2+\sigma_{\rm n}^2) \leq |h_{\rm p}|^2 \right\}\vert \ |h_{\rm sp}|^2 \right] \\
&=\int_0^{\infty} \! \frac{1}{\sigma_{\rm sp}}\exp{(-\frac{|h_{sp}|^2}{\sigma_{sp}})}. \exp{(\frac{-(2^{\mathcal{R}_{\rm p}}-1)}{P_{\rm p} \sigma_{\rm p}}.(\frac{E_t}{T} |h_{\rm sp}|^2+\sigma_{\rm n}^2))} \, \mathrm{d}|h_{\rm sp}|^2\\
& =\overline{P}^{\rm p}_{\rm nsu, out} \frac{P_{\rm p} \sigma_{\rm p}}{P_{\rm p} \sigma_{\rm p}+(\frac{E_t}{T}) \sigma_{\rm sp} (2^{\mathcal{R}_{\rm p}}-1)}  \cdot
\end{split}
\end{equation}
\end{small}
The transmission power of the SU in the sensing case is $E_t/T_t$, where $T_t=T- \tau$ is transmission time. Accordingly, the probability of no PU outage in case of the SU mis-detection $\overline{P}^{\rm p}_{\rm md, out}$ can be calculated as follows
\begin{small}
\begin{equation}
\begin{split}
& \overline{P}^{\rm p}_{\rm md, out}=\overline{P}^{\rm p}_{\rm nsu, out} \frac{P_p \sigma_{\rm p}}{P_{\rm p} \sigma_{\rm p}+(\frac{E_t}{T_t}) \sigma_{\rm sp} (2^{\mathcal{R}_{\rm p}}-1)},
\end{split}
\end{equation}
\end{small}Based on the previous analysis, $\mu_p$ can be expressed as
\begin{small}
\begin{align}
\label{PU Service Rate}
\begin{split}
& \mu_{\rm p}=\sum_{q_e=0}^{\mathcal{N}_t-1}\pi_{q_e}\!\overline{{P}}^{\rm p}_{\rm nsu, out} +\sum_{q_e=\mathcal{N}_t}^{\mathcal{N}_t+\mathcal{N}_s-1}\!\pi_{q_e}(\alpha_{q_e} \overline{{P}}^{\rm p}_{\rm ws, out}+\overline{\alpha}_{q_e} \overline{{P}}^{\rm p}_{\rm nsu, out})\\
&+\sum_{q_e=\mathcal{N}_t+N_s}^{N_{\rm max}}\pi_{q_e} (\beta_{1_{q_e}} \overline{{P}}^{\rm p}_{\rm ws, out} +\beta_{2_{q_e}} (P_D \overline{{P}}^{\rm p}_{\rm nsu, out}+P_{M} \overline{{P}}^{\rm p}_{\rm md, out})+\\&(1-\beta_{1_{q_e}}-\beta_{2_{q_e}}) \overline{{P}}^{\rm p}_{\rm nsu, out})\cdot
\end{split}
\end{align}
\end{small}where $\begin{small}{P_M=1-P_D}\end{small}$ is the average mis-detection probability. Define $\begin{small}{\mathcal{R}_s^s= b_{\rm s}/ T_t/ W }\end{small}$ and $\begin{small}{\mathcal{R}_s^{ws}= b_{\rm s}/ T/ W }\end{small}$ as the SU spectral efficiency in the sensing case and the case of access without sensing respectively. The probability of no SU outage in case of access without sensing and the PU is inactive is given by
\begin{small}
\begin{equation}
\begin{split}
& \overline{P}^s_{\rm{ws, out}}= 
 \exp \left( \frac{-(2^{\mathcal{R}_s^{ws}}-1) \sigma_n^2} {(E_t/ T)  \sigma_{s}}  \right)\cdot
\end{split}
\end{equation}
\end{small}Define $\overline{P}^s_{ \rm wsp, out}$ as the probability of no SU outage in
case of access without sensing and the PU is active
\begin{small}
\begin{equation}
  \overline{P}^s_{ \rm wsp, out} = 
 \overline{P}^s_{ \rm ws, out} \frac{(E_t/T) \sigma_{\rm s}}{(E_t/T) \sigma_{\rm s}+P_{\rm p} 
\sigma_{\rm ps} (2^{\mathcal{R}_s^{ws}}-1)}\cdot
\end{equation}
\end{small}Similarly, define $\overline{P}^s_{\rm s, out}$ as the probability of no SU outage when the PU is inactive and there is no false alarm
\begin{small}
\begin{equation}
  \overline{P}^s_{\rm s, out} = \exp \left( \frac{-(2^{\mathcal{R}_s^{s}}-1) \sigma_n^2} {(E_t/ T_t)  \sigma_{s}}  \right) \cdot
\end{equation}
\end{small}Define $\overline{P}^s_{\rm sp, out}$ as the probability of no SU outage in
the sensing case when the PU is active and mis-detected,
\begin{small}
\begin{equation}
 \overline{P}^s_{\rm sp, out} = \overline{P}^s_{\rm s, out} \frac{(E_t/T_t)\sigma_{\rm s}}{(E_t/T_t) \sigma_{\rm s}+P_{\rm p} 
\sigma_{\rm ps} (2^{\mathcal{R}_s^{s}}-1)} \cdot
\end{equation}
\end{small}
Based on the above definitions, the SU success rate defined as the probability of no SU outage can be expressed as follows
\begin{small}
\begin{equation}
\label{SU Service Rate}
\begin{split}
& \mu_s= \sum_{q_e=\mathcal{N}_t}^{\mathcal{N}_t+\mathcal{N}_s-1} \pi_{q_e} (\alpha_{q_e} (\rho \overline{P}^s_{\rm wsp, out} + \overline{\rho} \ \overline{P}^s_{\rm ws, out}))\\
& +\sum_{q_e=\mathcal{N}_t+N_s}^{N_{\rm max}} \pi_{q_e}(\beta_{1_{q_e}} (\rho \overline{P}^s_{\rm wsp, out}  + \overline{\rho} \ \overline{P}^s_{\rm ws, out})+ \\
& \beta_{2_{q_e}} (\rho P_{M} \overline{P}^s_{\rm sp, out}+\overline{\rho} \overline{P_F} \ \overline{P}^s_{\rm s, out}))\cdot
\end{split}
\end{equation}
\end{small}
Maximizing the success rate of the SU under PU QoS constraint ($\mu_{\rm th}$) is formulated as follows
\begin{small}
 \begin{equation}
\label{Optimization Problem }
\begin{aligned}
& \underset{\textbf{$\vect{\alpha}$, $\vect{\beta_{1}}$,$\vect{\beta_{2}}$, $\vect{\pi}$, $\tau$, $\lambda$}}{\text{max}}
& & \mu_s \\
& \text{subject to}
& &  \vect{0} \preceq \vect{\alpha},\vect{\beta_1}, \vect{\beta_2}, \vect{\pi}^{\rm t} \preceq \vect{1}\\
&& &\vect{\beta_1}+\vect{\beta_2} \preceq \vect{1} \\
&& & \vect{\pi} (P-\rm I)=0 , \  \vect{\pi} \vect{1} =1, \\
&& & \tau \in \{\tau_{\rm min}, 2 \tau_{\rm min}, ..., T-\tau_{\rm min} \}\\
&& & \lambda >0 \\
&& & \mu_p \geq \mu_{\rm th}
\end{aligned}
\end{equation}
\end{small}where $\begin{small}{\vect{\alpha}=[\alpha_{\mathcal{N}_t}, \dotsc,\alpha_{\mathcal{N}_t+\mathcal{N}_s-1}]^t, \      
 \vect{\beta_1}=[\beta_{1_{\mathcal{N}_t+\mathcal{N}_s}}, \dotsc,\beta_{1_{N_{\rm max}}}]^t}\end{small}$, $\begin{small}{\vect{\beta_2}=[\beta_{2_{\mathcal{N}_t+\mathcal{N}_s}}, \dotsc,\beta_{2_{N_{\rm max}}}]^t}\end{small}$, $\rm I$ is the identity matrix, $\mathbf{1}$ and $\mathbf{0}$ are column vectors whose entries are all equal to
$1$ and $0$ respectively and $\tau_{\rm min}$ is the minimum sensing time. We discretize $\lambda$ and adopt an exhaustive search approach over $\tau$ and $\lambda$ \cite{ewaisha2011optimization}. For a given $\tau$ and $\lambda$, we introduce an equivalent problem by making the following substitutions  
\begin{small}
\begin{eqnarray}
\tilde{\alpha}_{q_e}=\pi_{q_e}\alpha_{q_e},\ \ \ q_e=\mathcal{N}_t,\cdots,\mathcal{N}_t+\mathcal{N}_s-1\\
\tilde{\beta}_{1_{q_e}}=\pi_{q_e}\beta_{1_{q_e}},\ \ \ q_e=\mathcal{N}_t+\mathcal{N}_s,\cdots,N_{\rm{max}}\\
\tilde{\beta}_{2_{q_e}}=\pi_{q_e}\beta_{2_{q_e}},\ \ \ q_e=\mathcal{N}_t+\mathcal{N}_s,\cdots,N_{\rm{max}}
\end{eqnarray}
\end{small}Clearly, the problem now is a linear program in $\tilde{\alpha}_{q_e}$, $\tilde{\beta}_{1_{q_e}}$, $\tilde{\beta}_{2_{q_e}}$ and $\pi_{q_e}$. It can be solved efficiently using the software package\cite{grant2008cvx}. Finally, we consider the sensing only scheme in which the SU may sense the channel or remain silent and cannot access without the sensing, i.e, $\vect{\alpha}=\vect{0}$ and $\vect{\beta_1}=\vect{0}$.
 \section{Simulation Results}
\label{Simulation Results}
\begin{figure} [h]
  \centering
\includegraphics[width=1\linewidth,height=.12\textheight]{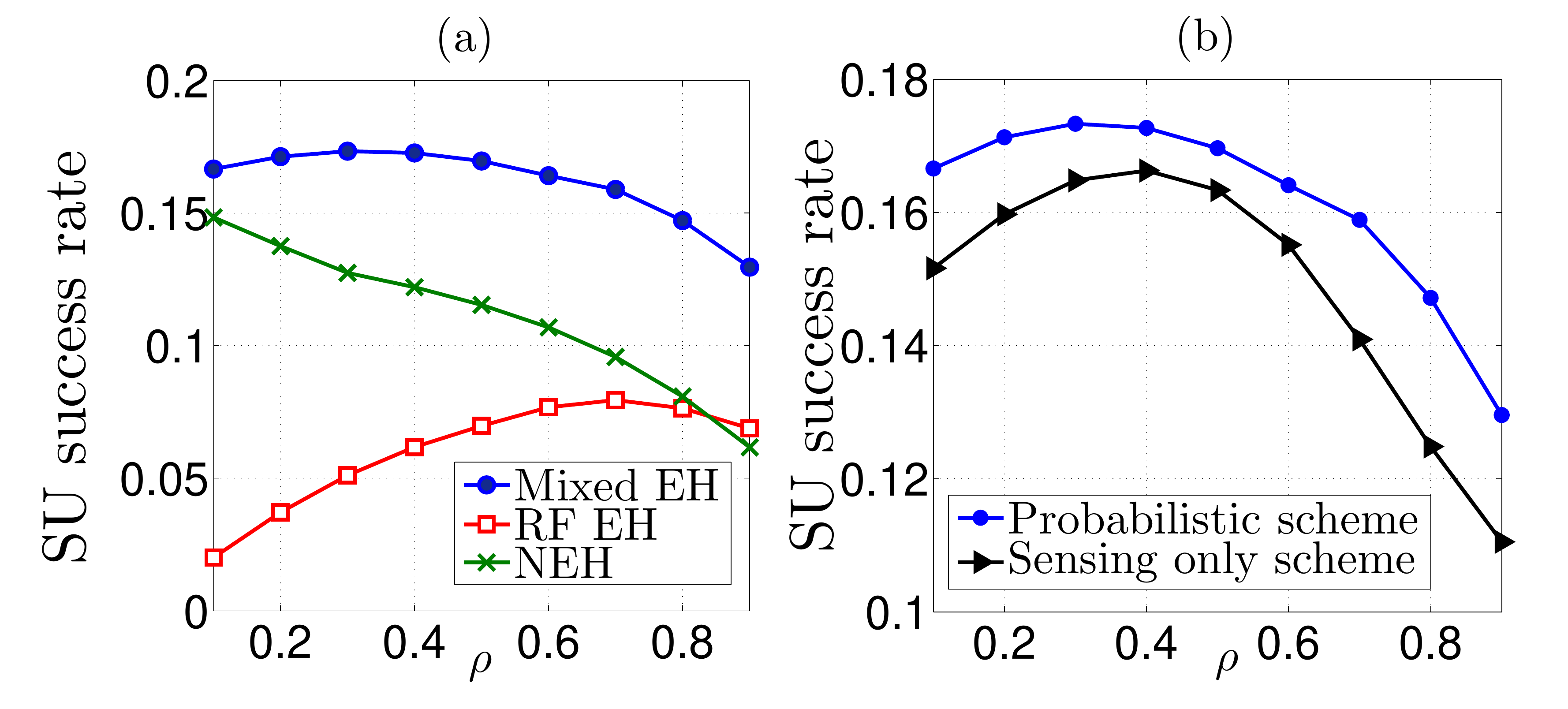} 
\caption{\small SU Success Rate\label{fig 2}}
\end{figure}

In this section, we present some numerical results which indicate the effectiveness of the mixed EH and the probabilistic scheme. The parameters that we have used in the simulation are summarized in Table. \ref{Simulation Parameters}. In Fig. \ref{fig 2} (a), $\mu_{\rm s}$ of the probabilistic scheme is shown. In the mixed EH and the RF EH cases, at low values of $\rho$, $\mu_{\rm s}$ increases as $\rho$ increases because the SU harvests more energy and can access the channel due to the PU's low activity. At high values of $\rho$, the SU harvests significant RF energy, but $\mu_s$ decreases due to the high activity of the PU. Intuitively, the mixed EH is better than the NEH and the RF EH as the SU in this case has much more energy.
\begin{table}
\caption{\small Simulation Parameters \label{Simulation Parameters}}
\centering
 \begin{tabular}{|c| c| c| c| c| c|} 
 \hline
 $\begin{small}{\mu_{\rm th}=0.65} \end{small}$ & $P_{\rm p}=4$ Watt & $b_{\rm p}=32$ bit & $b_{\rm s}=16$ bit  \\ \hline
 $\begin{small}{W=20} \end{small}$ \rm{KHz} & $\eta=0.5$ & $\sigma_{\rm n}^2=0.02$ \rm{Watt} & $T=1$ \rm{ms} \\ \hline
  $E_t=0.5$  \rm Joule &  $E_u=0.06$ \rm Joule & $e_{\rm proc}=10^{-2}$& $N_{\rm max}=20$ \\ \hline
  $\sigma_{\rm pst}=0.8/3^2$& $\sigma_{\rm ps}=0.8/5^2$ & $\sigma_{\rm sp}=0.8/5^2$ & $\sigma_{\rm p}=0.8/5^2$ \\ \hline
  $\sigma_{\rm s}=0.8/3^2$ &$\lambda_e=2 \ \rm{packets / sec}$ & $\mathcal{N}_t=9$ & \\ \hline
 \end{tabular}
\end{table} 
In Fig. \ref{fig 2} (b), we compare the probabilistic scheme with the sensing only scheme. As expected the probabilistic scheme outperforms the sensing only scheme. In the sensing only scheme, the SU has to sense or remain idle. At low values of $\rho$, the SU should exploit the opportunity and access the channel, but the energy is limited and therefore the SU may not be able to sense. At high values of $\rho$, the harvested RF energy is high and SU can sense. The channel is most likely to be occupied, hence sensing is not only waste of energy, but also decreases the SU's transmission time. 

\begin{figure} [h]
  \centering
\includegraphics[width=0.95\linewidth,height=.105\textheight]{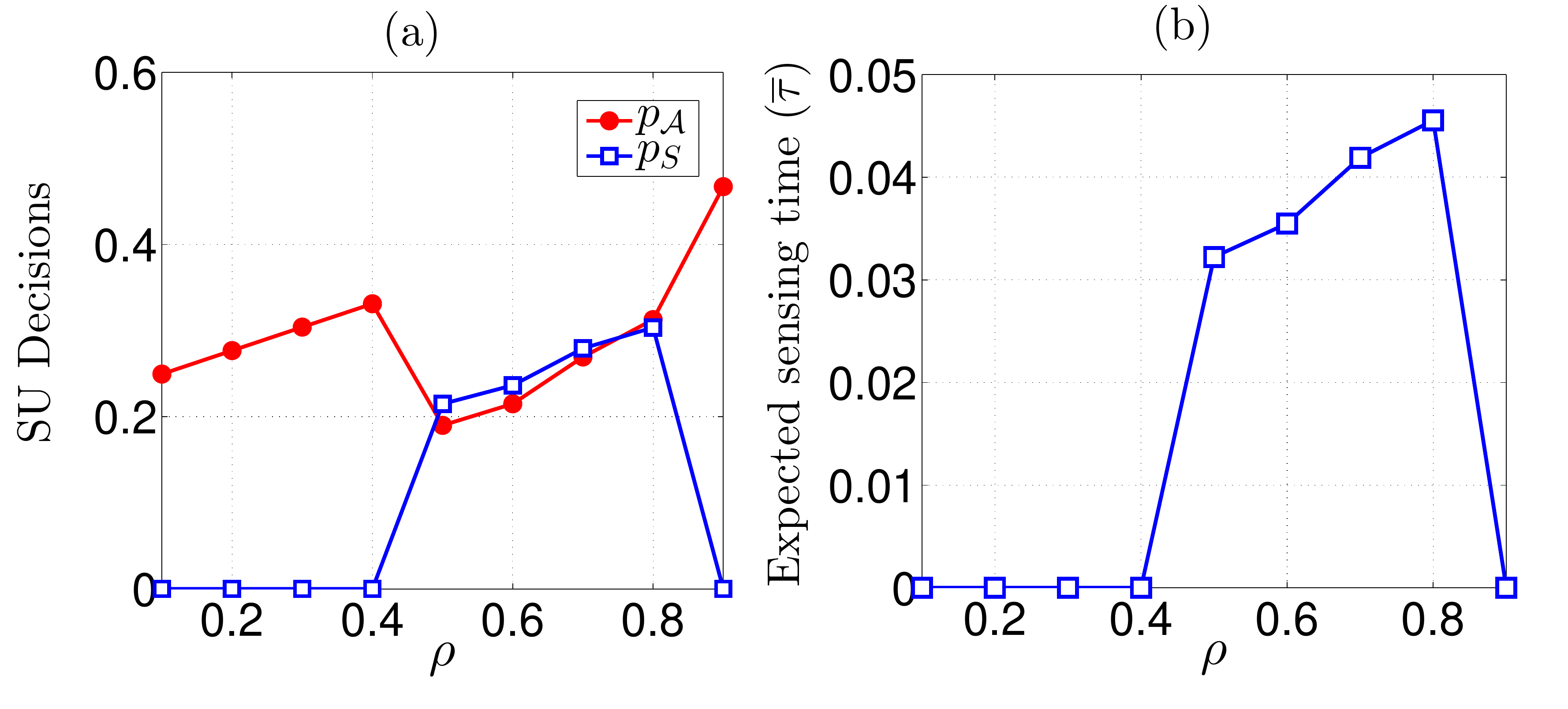} 
\caption{\small (a) Sensing and access without sensing probabilities, (b) The expected sensing time\label{fig 3}}
\end{figure}

\begin{figure} [h]
  \centering
\includegraphics[width=0.55\linewidth,height=.105\textheight]{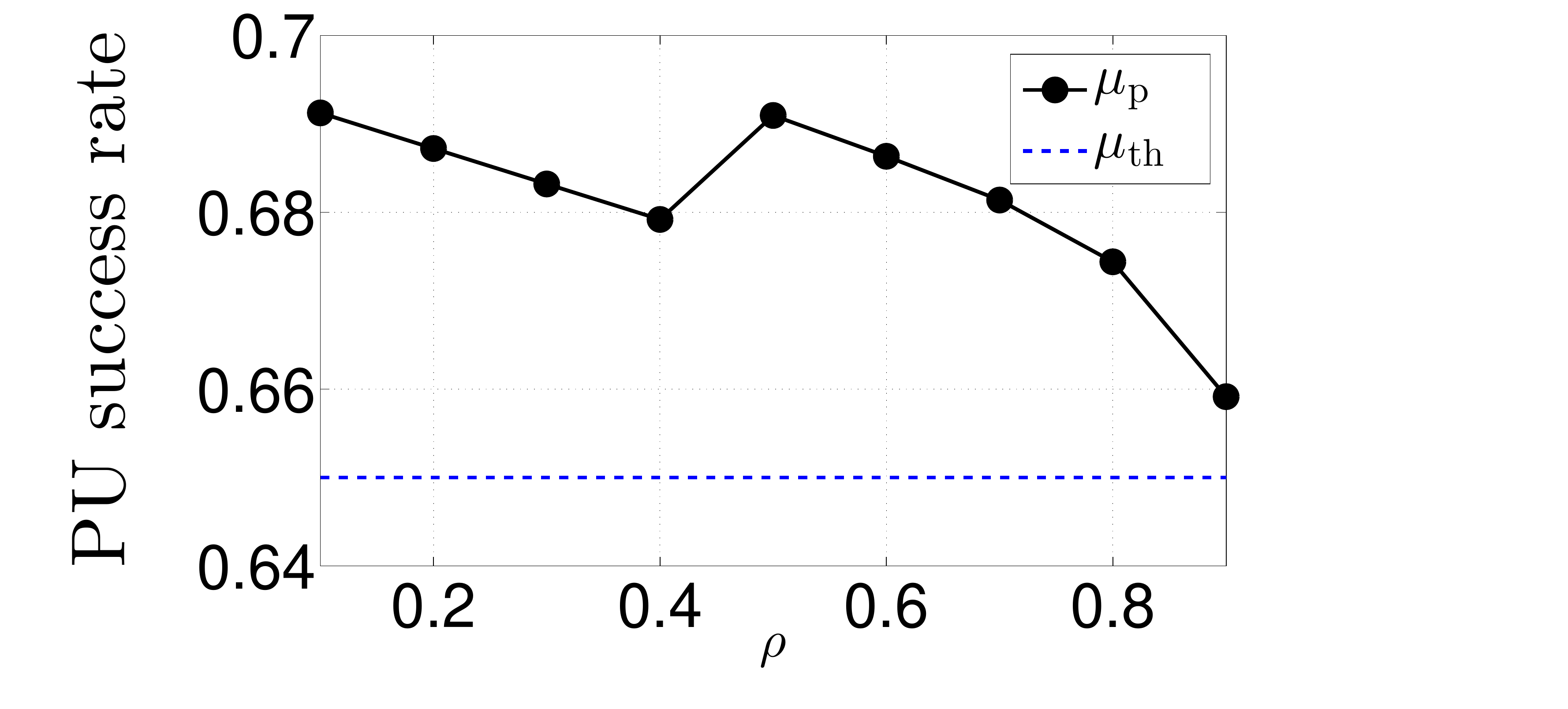} 
\caption{\small PU success rate \label{fig 4}}
\end{figure}

In Fig. \ref{fig 3}, $p_S$, $p_{\mathcal{A}}$ and $\overline{\tau}$ of the mixed EH probabilistic scheme are shown as a function of $\rho$. At low values of $\rho$, sensing is not useful as the channel is idle with high probability. The SU then prefers to access without sensing to save the sensing energy and to increase the transmission time. As $\rho$ increases, the SU harvests a significant RF energy. The SU then may sense the channel to avoid the PU interference. At extremely high values of $\rho$, the channel is occupied with high probability. Sensing in this case does not provide a new information for the SU. The SU then decides to access without sensing with high probability which is limited by the PU QoS constraint.
In Fig. \ref{fig 4}, $\mu_p$ in the mixed EH case probabilistic scheme is shown as a function of $\rho$. Interestingly, when the SU starts to sense the channel, $\mu_p$ increases because the sensing provides more primary protection. The SU then starts to be more aggressive as $\rho$ increases because it harvests significant RF energy and that causes $\mu_{\rm p}$ to decrease. 
\section{Conclusions}
\label{Conclusions}
In this paper, we have investigated a CR system with the mixed energy harvesting technique. We have proposed an energy-based probabilistic scheme in which the SU can opportunistically sense and access the PU's channel provided that the PU's QoS is satisfied. The decision is based on the available energy and the PU's activity. We have concluded that sensing the spectrum is not always useful as the SU may prefer to access without sensing to save the sensing energy and dedicate more time for transmission.
\vspace{-8 pt}
\section*{Appendix}
\label{Appendix}
The probability of harvesting $q$ packets from both nature and RF signals $p_Q(q)$ can be calculated as follows
\begin{small}
\begin{equation}
p_Q(q)=\sum_{n+r=q} p_N(n) p_{\mathcal{R}_e}( r).
\end{equation}
\end{small}
Following the description in Section \ref{Sensing}, the elements of $P$ can be calculated as follows,
\begin{enumerate}
\item $j < N_{\rm max}$ 
\begin{itemize}
\item $i < \mathcal{N}_t$
\vspace{-10 pt}
\begin{small}
\begin{equation}
P_{ij}=\overline{\rho} \ p_N(j-i)+ \rho \ p_Q(j-i) .
\end{equation}
\end{small}
\item $\mathcal{N}_t \leq i < \mathcal{N}_t + \mathcal{N}_s$
\begin{small}
\begin{equation}
\begin{split}
& P_{ij}=\overline{\rho} \alpha_{i} p_N(j-i+\mathcal{N}_t)+\overline{\rho} \ \overline{\alpha}_{i} p_N(j-i)\\
& + \rho \ \alpha_{i} \ p_Q(j-i+\mathcal{N}_t)  + \rho\overline{\alpha}_{i} \ p_Q(j-i)  .
\end{split}
\end{equation}
\end{small}
\item $\mathcal{N}_t + \mathcal{N}_s \leq i \leq N_{\rm max} $
\begin{small}
\begin{equation}
\begin{split}
& P_{ij}= \overline{\rho} \beta_{1_{i}} p_N(j-i+\mathcal{N}_t)+ \overline{\rho}  \beta_{2_{i}} P_F  p_N(j-i+\mathcal{N}_s)+\\
& \overline{\rho}  \beta_{2_{i}}  \overline{P_F}  p_N(j-i+\mathcal{N}_s+\mathcal{N}_t)+ \overline{\rho}  (1-\beta_{1_{i}}-\beta_{2_{i}})  p_N(j-i)+\\
&\rho \beta_{1_{i}} p_Q(j-i+\mathcal{N}_t)+\rho  \beta_{2_{i}}P_D  p_Q(j-i+\mathcal{N}_s)+\\
&\rho  \beta_{1_{i}} P_{M}p_Q(j-i+\mathcal{N}_s+\mathcal{N}_t)
+\!\rho(1-\beta_{1_{i}}-\beta_{2_{i}})p_Q(j-i) .
\end{split}
\end{equation}
\end{small}
\end{itemize}

\item $j = N_{\rm max}$

The transition probabilities in this case have different formulas due to the battery limited capacity. They can be derived from the above equations by replacing $p_N(n)$ by $\Pr (N \geq n)$  and replacing $p_Q(q)$ by $\Pr (Q \geq q)$.
\end{enumerate}
\balance
\bibliographystyle
{IEEEtran}
\bibliography{IEEEabrv,Nulls}

\end{document}